\documentclass[%
reprint,
superscriptaddress,
amsmath,
amssymb,
aps,
]{revtex4-2}

\usepackage{CJK}
\usepackage{graphicx}
\usepackage{epstopdf}
\usepackage{dcolumn}
\usepackage{bm}

\usepackage{flushend}

\usepackage
[colorlinks,
linkcolor=blue,
anchorcolor=blue,
urlcolor=blue,
citecolor=blue
]{hyperref}

\usepackage{color}
\usepackage{upgreek}
\usepackage{makecell}

\usepackage{textcomp}

\usepackage{threeparttable}

\usepackage{gensymb}

\usepackage{hyperref}

\usepackage{subfigure} 
\usepackage[utf8]{inputenc}
\usepackage[T1]{fontenc}
\usepackage{mathptmx}
\usepackage{etoolbox}
\DeclareUnicodeCharacter{202F}{\,}
\usepackage{CJK}
\usepackage{epstopdf}
\usepackage{float} 
\usepackage{siunitx} 
\usepackage[T1]{fontenc} 
\usepackage{textcomp} 
\usepackage{amsmath} 
\usepackage{color}
\usepackage{upgreek}
\usepackage{makecell}
\usepackage{textcomp}
\usepackage{threeparttable}
\usepackage{gensymb}

\begin{document}
	
	\preprint{APS/123-QED}
	
	\title{Effect of the ${\rm^{15}N(p,\alpha)^{12}C}$ reaction on the kinetic energy release of water molecule fragmentation}
	
	\author{Zhuohang He}
	\thanks{These authors contribute equally to this work.}
	\affiliation{Key Laboratory of Radiation Physics and Technology of Ministry of Education, Institute of Nuclear Science and Technology, Sichuan University, Chengdu 610064, China}
	
	\author{Zhencen He}
	\thanks{These authors contribute equally to this work.}
	\affiliation{West China School of Basic Medical Sciences and Forensic Medicine, Sichuan University, Chengdu 610041, China}
	\affiliation{Key Laboratory of Radiation Physics and Technology of Ministry of Education, Institute of Nuclear Science and Technology, Sichuan University, Chengdu 610064, China}
	
	\author{Mingliang Duan}
	\affiliation{Key Laboratory of Radiation Physics and Technology of Ministry of Education, Institute of Nuclear Science and Technology, Sichuan University, Chengdu 610064, China}
	
	\author{Junxiang Wu}
	\affiliation{Key Laboratory of Radiation Physics and Technology of Ministry of Education, Institute of Nuclear Science and Technology, Sichuan University, Chengdu 610064, China}
	\affiliation{Radiation Oncology Department Sichuan Cancer Hospital and Institute, Affiliated Cancer Hospital of University of Electronic Science and Technology of China, Chengdu 610041, China}

	\author{Liyuan Deng}
	\affiliation{Key Laboratory of Radiation Physics and Technology of Ministry of Education, Institute of Nuclear Science and Technology, Sichuan University, Chengdu 610064, China}
	
	\author{Ziqi Chen}
	\affiliation{Key Laboratory of Radiation Physics and Technology of Ministry of Education, Institute of Nuclear Science and Technology, Sichuan University, Chengdu 610064, China}
	
	\author{Shuyu Zhang}
	\affiliation{West China School of Basic Medical Sciences and Forensic Medicine, Sichuan University, Chengdu 610041, China}
	
	\author{Zhimin Hu}
	\email{huzhimin@scu.edu.cn}
	\affiliation{Key Laboratory of Radiation Physics and Technology of Ministry of Education, Institute of Nuclear Science and Technology, Sichuan University, Chengdu 610064, China}
	
	\date{\today}
	
	\begin{abstract}
		
		In this work, we investigated the effect of ${\rm^{15}N(p,\alpha)^{12}C}$ reaction produced by the collision between proton and ammonia monohydrate on the kinetic energy release (KER) of water molecule fragmentation. After the occurrence of the nuclear reaction, it was found that the charge states $q$ and the flight speeds $v$ are the main factors affecting the KER of water molecule fragmentation. With the value of $q/v$ increases, the KER distribution gets wider and the peak position changes more pronounced. The energy gained by each fragment is related to the mass of the fragment and the distance of the fragment from the nuclear reaction. In this study, the fragments with smaller masses and the distances far away from the nuclear reaction get higher energies. The fragments of water molecules getting higher energy may induce other factors affecting the radiotherapy effect, which needs more detailed investigations in the future.
		
	\end{abstract}
	
	\maketitle
	
	
	Radiation therapy (RT) is an effective clinical treatment method for malignant tumors and has been around for over one hundred years. In 1896, six months after W. Roentgen discovered X-rays, V. Despeignes first used X-rays for experimental treatment of cancer patients for the first time \cite{Despeignes1896}. In 1898, after Mr. and Mrs. Curie discovered the radioactive element radium, it was quickly used for internal irradiation of tumors \cite{pouget2015}. Currently, RT is used to treat most types of tumors, and approximately 50\% of cancer patients will receive RT during illness. It is estimated that RT's contribution to curative treatment is around 40\% \cite{baskar2012}. RT uses many types of rays, such as X-rays, electron beams, proton beams, neutron beams, and ion beams, to kill cancer cells by ionizing radiation effect \cite{Semwal2020}. Any radiation beams can destroy any cancer cells or living entities if the dose is high enough. However, the key factor in the development of radiation medicine has always been how to avoid irradiating normal tissue. In other words, to protect normal tissue while avoiding side effects that affect the treatment \cite{slater2012}. Due to the Bragg peak, ion beam therapy significantly reduces off-target exposures and protects normal tissues. Additionally, ion beams have stronger relative biological effectiveness (RBE) and better RT efficacy than other radiation beams.
	
	Boron neutron capture therapy (BNCT) makes good use of the strong RBE of heavy ions. The principle is to inject $^{10}$B-containing targeted drugs into the human body in advance, allowing the $^{10}$B to accumulate in the tumor tissue specifically. Then externally irradiate the tumor site through directional low-energy epithermal or thermal neutron beams, the ${\rm^{10}B(n,\alpha)^{7}Li}$ reaction is induced to generate alpha ($\alpha$) particles and lithium ions inside the tumor cells, and the particles with the nuclear-reaction energy kill the tumor cells. BNCT maximizes the protection of normal tissues while ensuring efficacy and has been clinically studied at several tumor sites with positive results \cite{miyatake2020, kawabata2021}. However, the complexity of the epithermal or thermal neutron beam generation process in BNCT programs significantly increases the cost of treatment. Although there are treatment options that use accelerators instead of nuclear reactors \cite{Suzuki2020}, the cost of the related equipment remains high, significantly impeding this technology's diffusion and application. In 2014, Yoon \textit{et al.} proposed an innovative approach known as proton boron capture therapy (PBCT) \cite{yoon2014,yoon2019}. PBCT utilizes protons to induce the reaction p+$^{11}$B→3$\alpha$ to produce high-energy particles. This approach greatly simplifies equipment requirements and reduces treatment costs. However, there are many problems with this solution in practice \cite{mazzone2019, Hosobuchi2023}. Water molecules are ubiquitous in the cellular environment. Therefore, high-energy particles from nuclear reactions tend to ionize these molecules, leading to the molecular fragmentation. In addition, subsequent nuclear reactions may affect the therapeutic efficacy by influencing the molecular fragmentation process. Therefore, it is important to study the effect of nuclear reactions on molecular fragmentation for clinical RT research.
	\begin{figure*}
		\centering
		\subfigure{
			\label{fig:1(a)}
			\includegraphics[width=0.3823\textwidth]{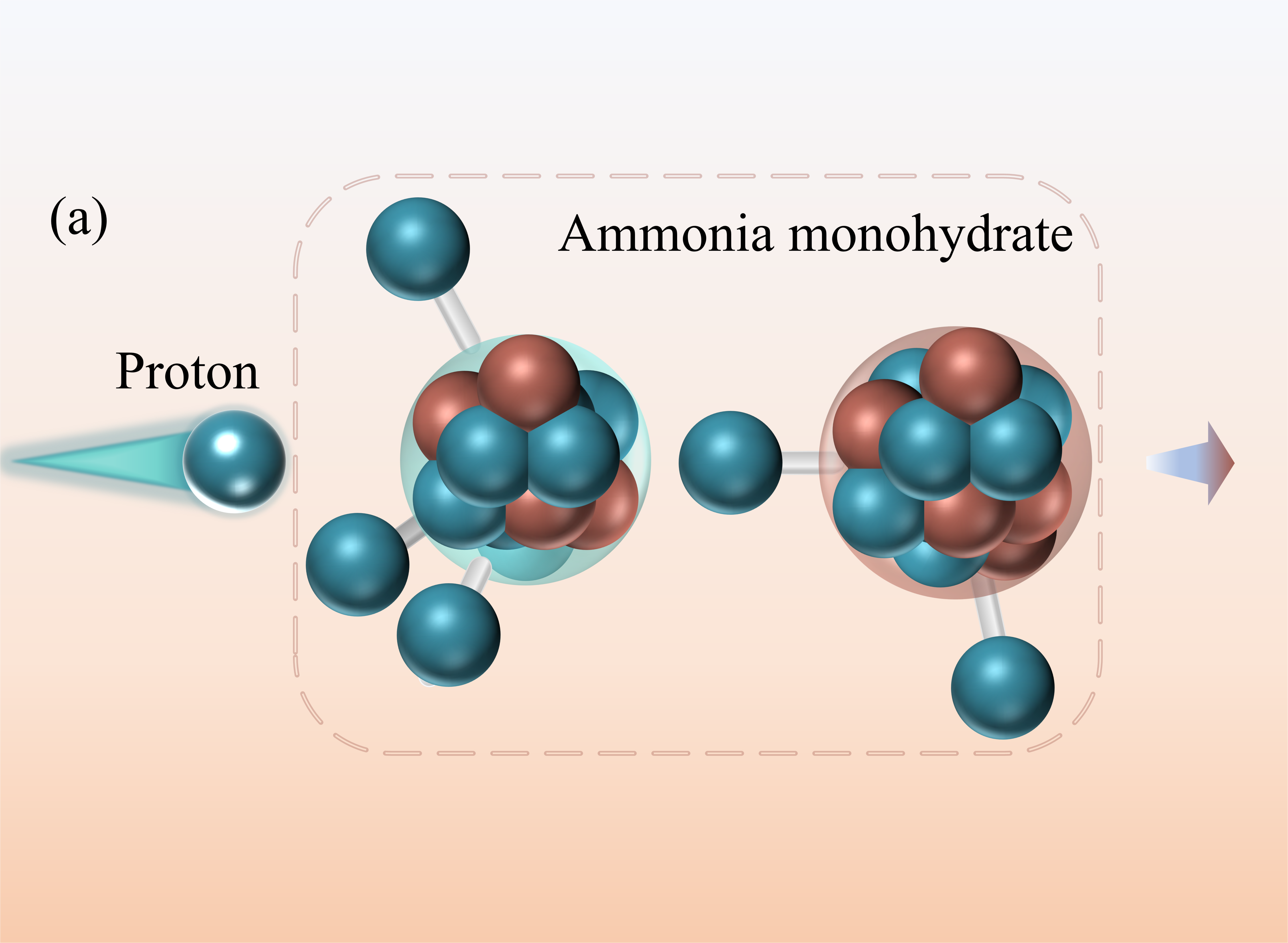}
			\hspace{-4mm}
		}
		\vspace{-0mm}
		\subfigure{
			\label{fig:1(b)}
			\includegraphics[width=0.3075\textwidth]{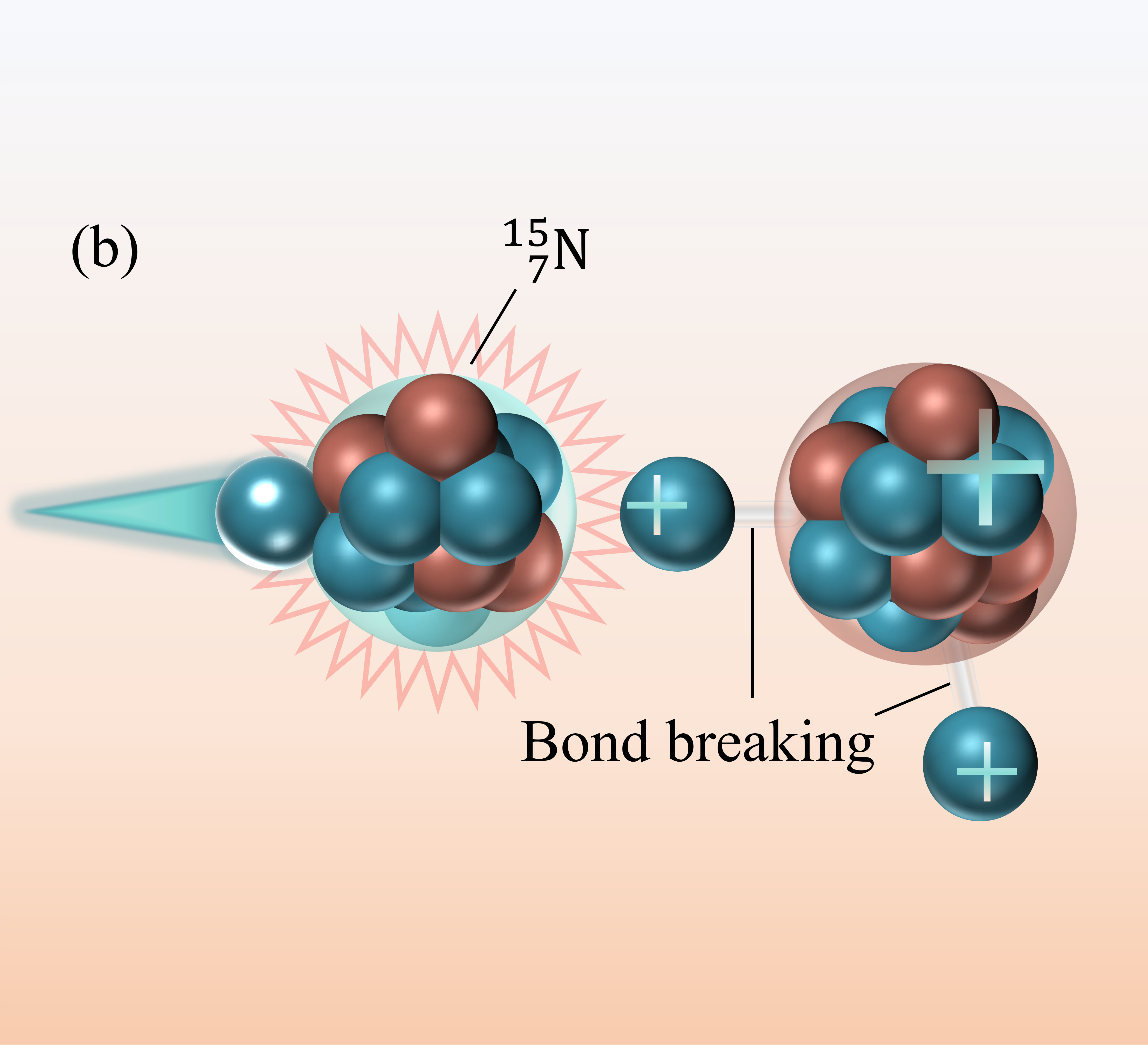}
			\hspace{-4mm}
		}
		\subfigure{
			\label{fig:1(c)}
			\includegraphics[width=0.28\textwidth]{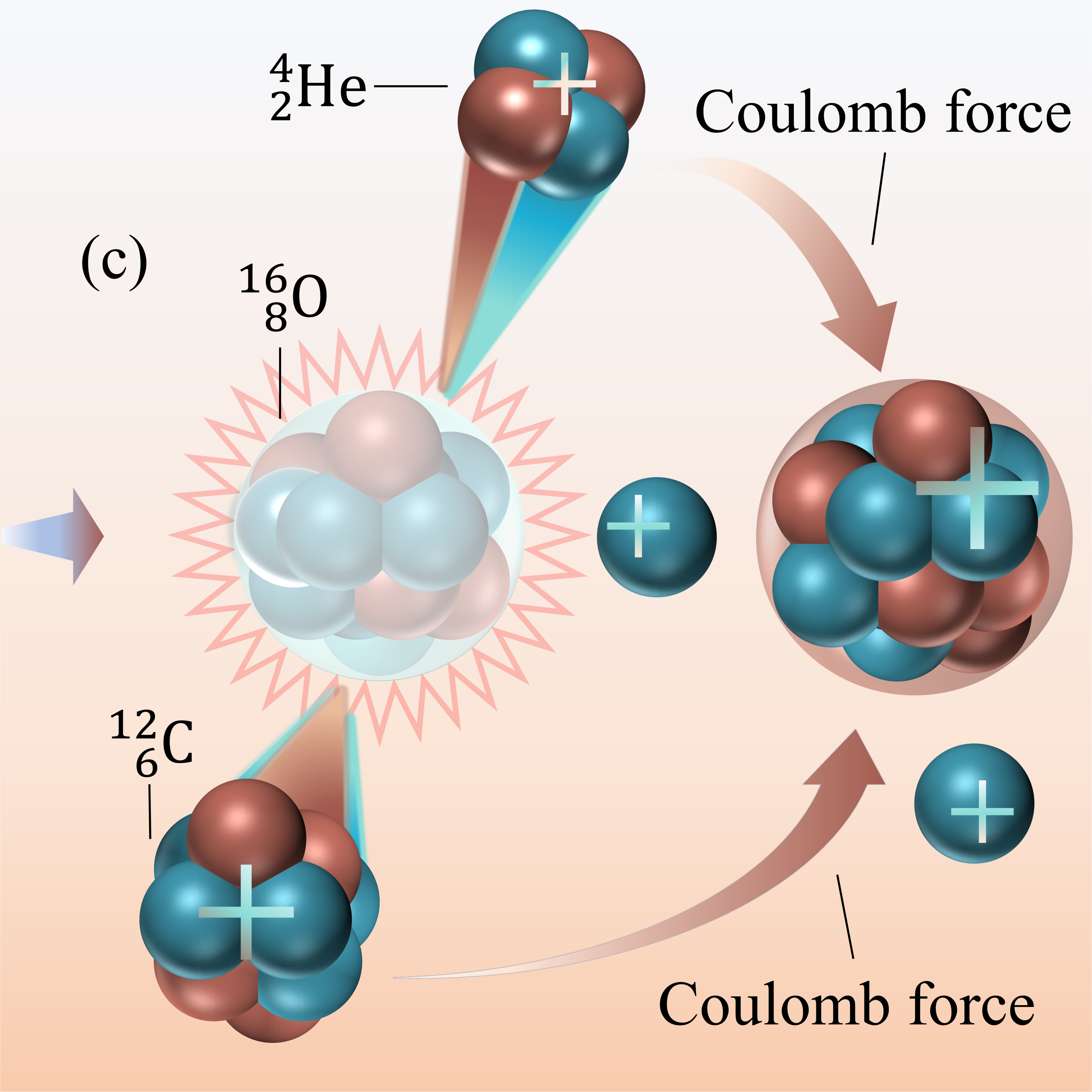}
			\hspace{0mm}
		}
		\vspace{-2mm}
		\caption{\label{fig:1}The ${\rm^{15}N(p,\alpha)^{12}C}$ reaction affects the water molecule fragmentation reaction process. (a) ${_{1}^{1}\rm{p}+^{15}\rm{NH}_{3}\cdot\rm{H_{2}O}}$ collision system. (b) Proton induced ${\rm^{15}N(p,\alpha)^{12}C}$ reaction while water molecule break bonds. (c) Particles emitted from the ${\rm^{15}N(p,\alpha)^{12}C}$ reaction affects water molecule fragmentation.}
		\vspace{-3.5mm}
	\end{figure*}
	
	Cold target recoil ion momentum spectroscopy (COLTRIMS) is a well-established technique to study the fragmentation dynamics of molecular cations. By analyzing the time-of-flights and positions of the collision products by a position-sensitive detector, the three-dimensional momentum vectors and kinetic energies (KEs) of the collision products are obtained \cite{dorner2000}. With the aid of COLTRIMS technology, many studies have focused on the molecular fragmentation mechanism in recent years \cite{He2022, Yan2023, Titze2011, khan2017, wang2020, ren2022, Li2019, Wei2021, wang2020water}. COLTRIMS can also used to study radiobiological processes. The Max-Planck-Institut has reported the process of intermolecular Coulombic decay of tetrahydrofuran-water dimers under electron-impact-induced ionization \cite{wang2020water, ren2018}. This is important for researching radiation damage and RT. However, studies related to nuclear reactions by COLTRIMS have not been reported. 
	
	In this Letter, we constructed a model based on the point charge approximation to simulate the effect of the ${\rm^{15}N(p,\alpha)^{12}C}$ reaction on the KER of water molecules. It observes that nuclear reactions broaden the KER, which may affect the energy deposition during the RT process or lead to other effects. The simulation results also offer references for subsequent experiments on the microscale mechanisms of radiation damage by using the COLTRIMS.
	
	The collision between proton and ammonia monohydrate (${^{15}\rm{NH}_{3}\cdot\rm{H_{2}O}}$) was chosen for this study. The reasons are as follows. Glutamine has recently gained attention as a targeted drug \cite{Reinfeld2021, Halama2022, Ma2022} because tumor cells depend heavily on this amino acid for growth and proliferation. Tumor cells have a high glutamine uptake, resulting in higher glutamine concentrations in tumor cells than in normal cells \cite{Reinfeld2021}. If this feature can be exploited, glutamine is allowed to accumulate in the tumor cells. Then irradiate the tumor cells with proton beams, which induces a ${\rm^{15}N(p,\alpha)^{12}C}$ reaction (reaction cross sections comparable to those in PBCT) that produces high-energy particles to kill the cancer cells \cite{schardt1952, Hosobuchi2023}. It produces a better RT effect because nitrogen is more likely to be enriched in the cancer cells and produces carbon ions and $\alpha$ particles with strong RBE \cite{wu2024}. Linear ammonia monohydrate [see Fig. \ref{fig:1(a)}] was chosen as a target because it has a simpler structure than glutamine and contains amino groups like glutamine. Additionally, ammonia monohydrate contains water molecules, making it easier to study how the ${\rm^{15}N(p,\alpha)^{12}C}$ reaction affects the water molecule fragmentation. COLTRIMS specializes in studying kinetic energy release (KER) during molecular fragmentation, so this study focuses on the KER change of water molecules.
	
	This study presents a model constructed on the point charge (PC) approximation. It is widely used for describing the Coulomb explosion (CE) process in molecular fragmentation. Under this approximation, the ionic fragments are regarded as point-charged particles and the mass and charge of each particle are located at the center of the nuclei. Meanwhile, the distance between these points is defined as the equilibrium internuclear distances of the neutral molecule \cite{wang2020}. The initial distance between particles is defined as the bond length. The motion of each particle is affected by the Coulomb repulsion from other particles. Here is an example of a KER calculation for a one-dimensional molecule. Coulomb's law is given by
	\begin{equation}
		\label{eq1}
		F_{i}=k \sum_{j \neq i} \frac{q_{i} q_{j}}{\left(x_{i}(t)-x_{j}(t)\right)^{2}}, 
	\end{equation}
	where $k$ represents the Coulomb's constant, $j$ represents particles other than $i$, and $q_{i}$ and $q_{j}$ stand for the charges of particles $i$ and $j$, the initial velocity of the particles can be written as
	\begin{equation}
		\label{eq2}
		\dot{x}_{i}(0)=0.
	\end{equation}
	The KE for each particle concerning propagation time is given by
	\begin{equation}
		\label{eq3}
		{\rm KE}_{i}(t)=\frac{1}{2} m_{i}\left(\dot{x}_{i}(t)\right)^{2}.
	\end{equation}
	KER is defined as the total KE of all fragment ions
	\begin{equation}
		\label{eq4}
		{\rm KER}=\sum_{i} {\rm KE}_{i}.
	\end{equation}
	
	According to the above equation, each molecular fragment is considered as a point charge, and finally the KER is obtained from the KE of each point charge.
	\begin{figure*}
		\centering
		\subfigure{
			\label{fig:2(a)}
			\includegraphics[width=0.49\textwidth]{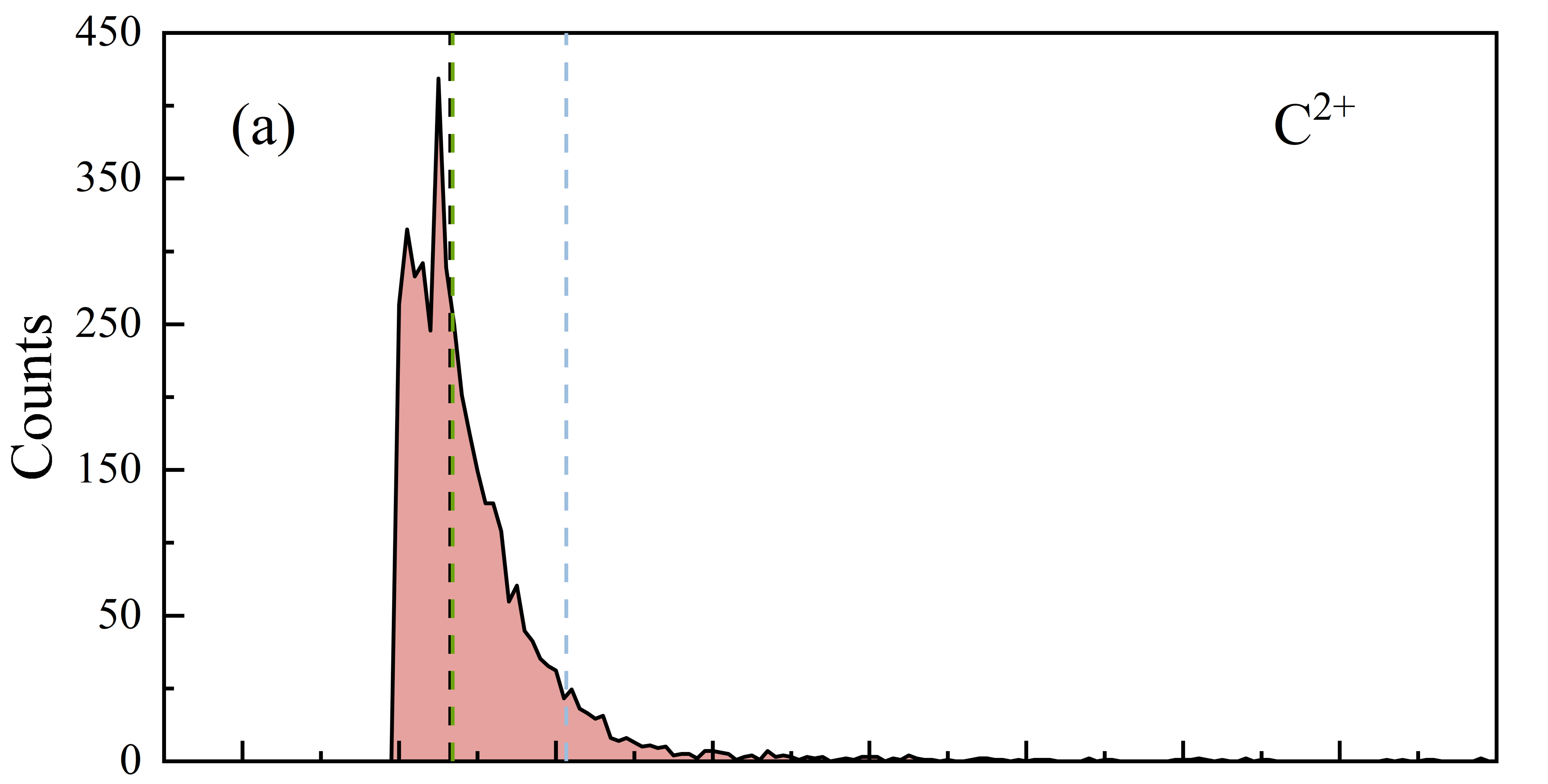}
			\hspace{-3mm}
		}
		\vspace{-3.5mm}
		\subfigure{
			\label{fig:2(b)}
			\includegraphics[width=0.49\textwidth]{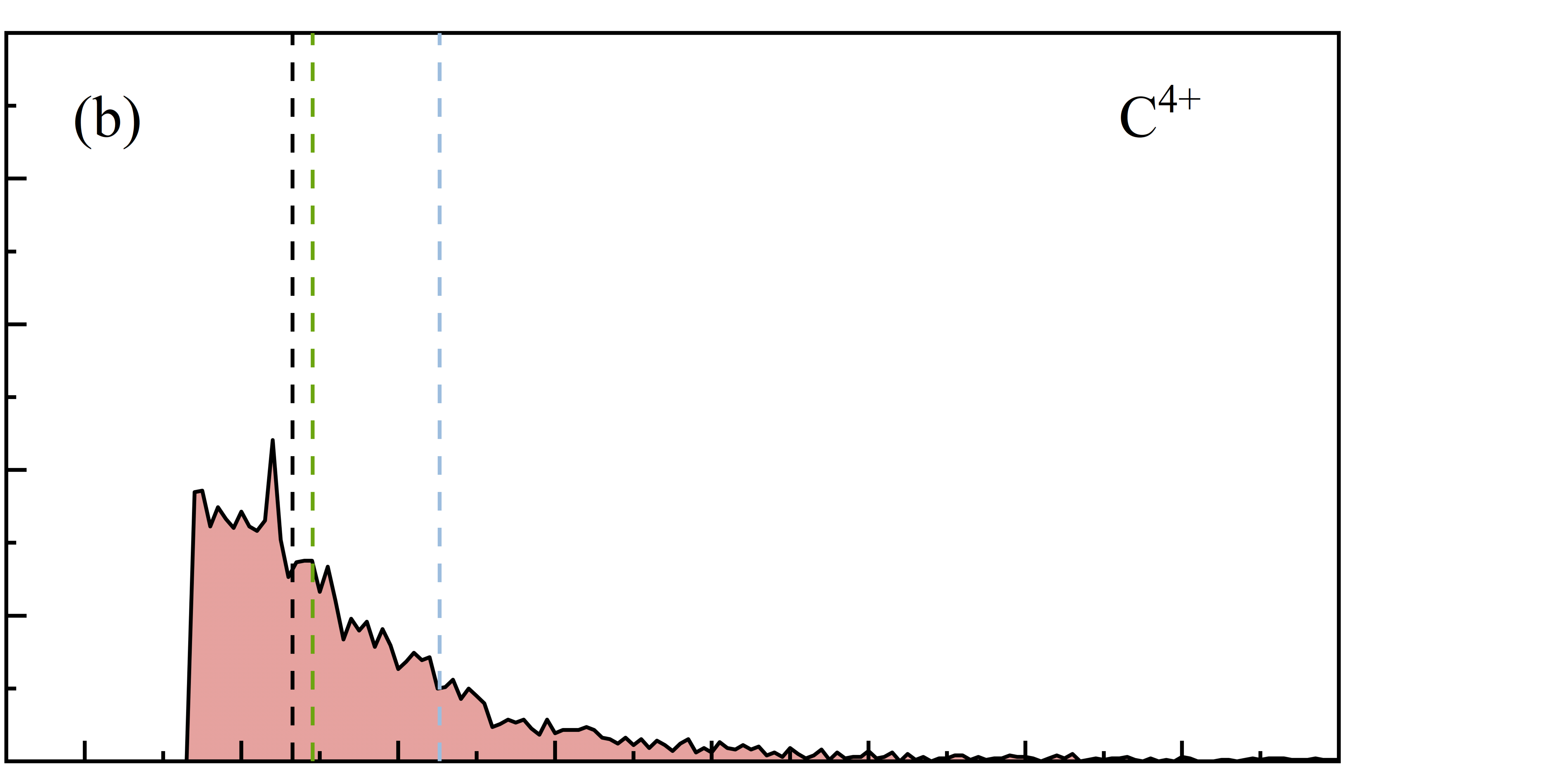}
			\hspace{0mm}
		}
		\subfigure{
			\label{fig:2(c)}
			\includegraphics[width=0.49\textwidth]{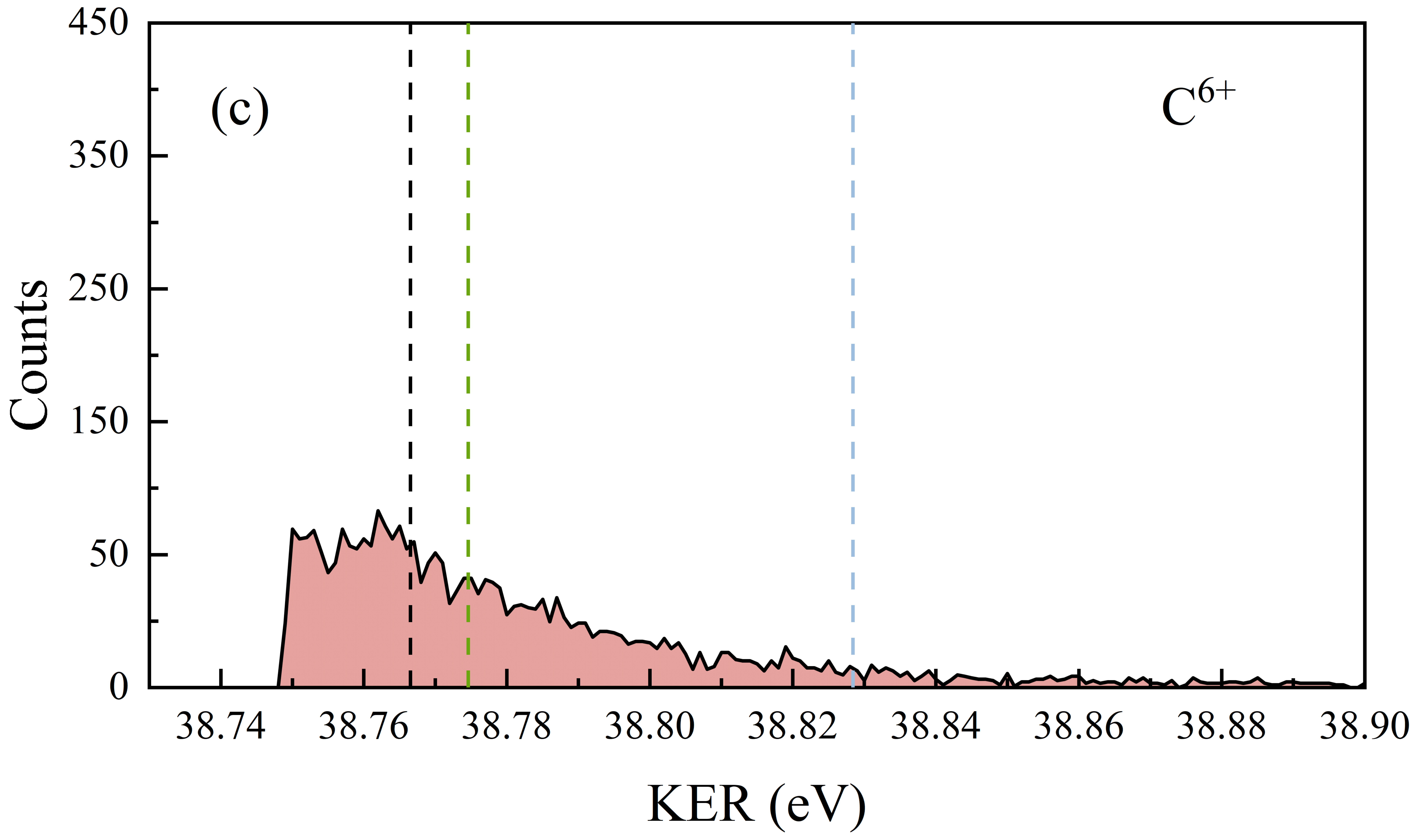}
			\hspace{-3mm}
		}
		\subfigure{
			\label{fig:2(d)}
			\includegraphics[width=0.49\textwidth]{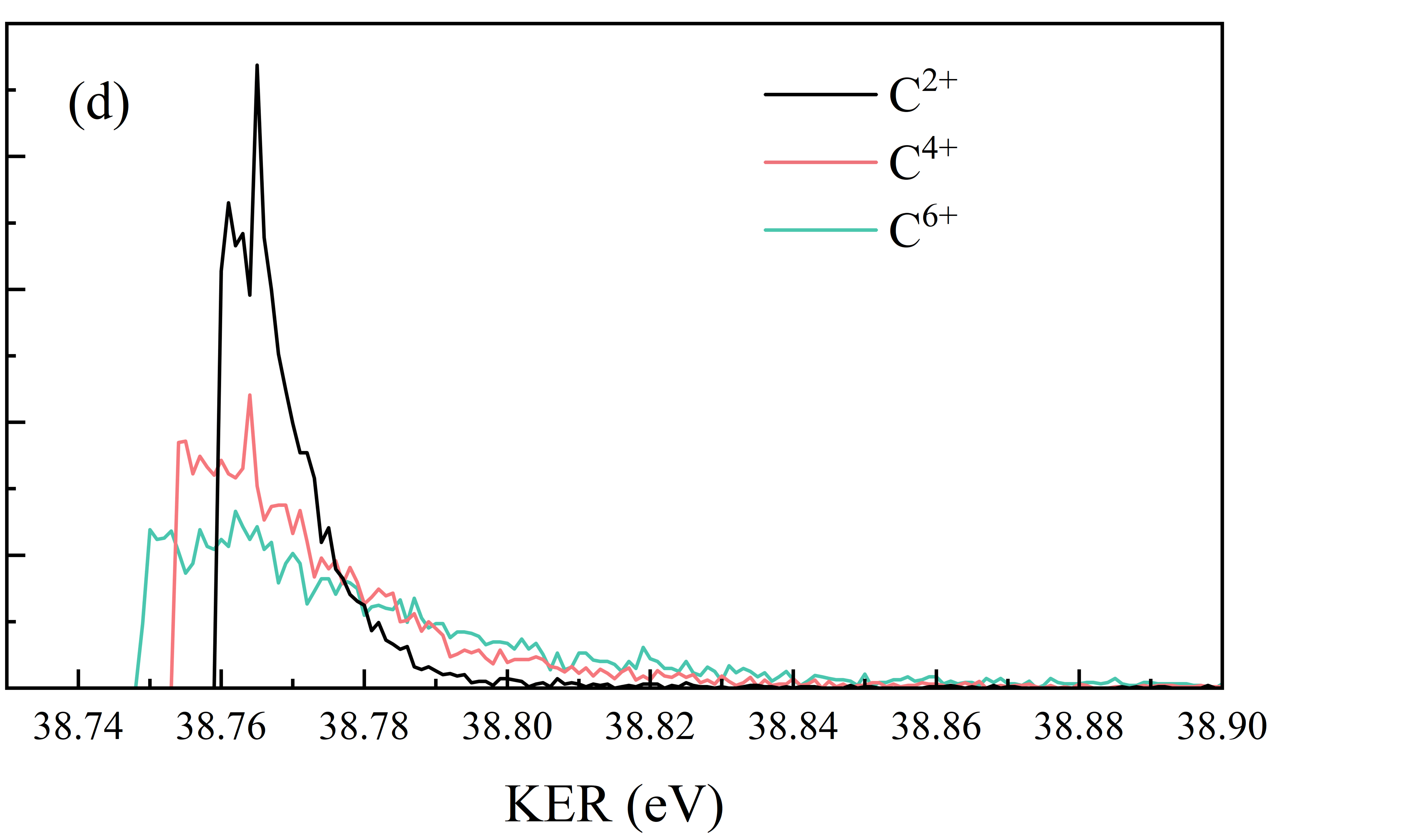}
		}
		\vspace{-2mm}
		\caption{\label{fig:2}KER distribution of water molecule fragmentation due to differently charged carbon ions from nuclear reactions. The black dotted line represents the KER from the effects of non-nuclear reactions, the green dotted line represents the median KER of water molecules, and the blue dotted line represents the average KER of water molecules.}
		\vspace{-4.5mm}
	\end{figure*}
	
	To investigate the effects of nuclear reactions on the molecular KER, it is necessary to couple nuclear reactions into the CE process. Bohr's compound nuclear model is used in this study to describe nuclear reactions \cite{mahaux1979}. The compound nuclear reaction can be divided into a completely independent two-step process. The first step involves the fusion of the incident particle with the target nucleus to form a compound nucleus. The subsequent step is the decay of the compound nucleus in an excited state. The compound nucleus reaction is given by
	\begin{equation}
		\label{eq5}
		\rm{a+A} \rightarrow \rm{C^*} \rightarrow \rm{B+b}.
	\end{equation}
	Here, ${\rm C^*}$ represents the compound nucleus in its excited state, with a lifetime ranging from 10$^{-18}$ s to 10$^{-14}$ s. After decay, the angular distribution of emitted particles is isotropic or axially symmetric \cite{mahaux1979}.
	
	We are assuming that the decay of the compound nucleus and the molecular fragmentation occurs at the same time [Figs. \ref{fig:1(b)} and \ref{fig:1(c)}]. The compound nuclear decay produces charged particles, and all the nuclear-reaction energy is converted into the KEs of charged particles. The particles produced by the nuclear reaction are regarded as a point charge. The nuclear reaction effect process reduces to Coulomb interactions between the nuclear-reaction products and the molecular fragments [Fig. \ref{fig:1(c)}].
	
	Before conducting simulations, it is necessary to define the nuclear reaction channel, molecular fragmentation channel, and molecular structure. Firstly, the 3 MV Tandetron accelerator in Sichuan University can accelerate projectiles to energy \textit{E} = (\textit{q}+1) $\times$ (0.2$\sim$3.0) MeV \cite{han2018}, where \textit{q} is the charge state of the ion projectiles. Thus, the accelerator can provide the proton with the energy of 0.4$\sim$6.0 MeV, exactly the residual energy of the proton at the Bragg peak position \cite{mazzone2019}. It is possible to simulate the reaction of protons with $\rm{^{15}N}$ in tumor cells. The proton reacts with $\rm{^{15}N}$ in this energy range with three main reaction channels, as follows:
	\begin{align}
		\label{eq6}
		\mathrm{^{15}}\rm{N}+\mathrm{^{1}}\rm{H} \rightarrow\left(\mathrm{^{16}}\rm{O}\right) & \rightarrow \mathrm{^{12}}\rm{C}+\mathrm{^{4}}\rm{He} & & Q  = 5.0\ \mathrm{MeV} \\
		& \rightarrow \mathrm{^{12}}\rm{C}^{*}+\mathrm{^{4}}\rm{He} & & Q  = 0.5\ \mathrm{MeV} \\
		& \rightarrow \mathrm{^{16}}\rm{O}+\gamma & & Q  = 12.1\ \mathrm{MeV}.
	\end{align}
	The focus is on channel (6) because it has the largest nuclear-reaction cross section \cite{schardt1952}. The energy $Q$ is completely transformed into the KE of the $\alpha$ particle and carbon ion. The motion of two particles may affect the process of water molecule fragmentation [see Fig. \ref{fig:1(c)}].
	\begin{equation}
		\label{eq9}
		\mathrm{H}_2 \mathrm{O}^{3+} \rightarrow \mathrm{H}^{+}+\mathrm{O}^{+}+\mathrm{H}^{+}.
	\end{equation}
	
	Water molecule fragmentation is divided into two ways: concerted and sequential processes \cite{Singh2013}. Since the PC model is effective in modeling concerted fragmentation, we chose the path described by Eq. (\ref{eq9}). The water molecule is defined with a bond length of 0.96 $\text{\AA}$ and a bond angle of 104.5$^{\circ}$ \cite{kim2016}.
	
	To compare with the effect of nuclear reaction, it is necessary to calculate the KER of Eq. (\ref{eq9}) without nuclear reaction influence. Using the previously established PC model, the KER is calculated to be about 38.77 eV. Werner \textit{et al.} \cite{werner1995} and Alvarado \textit{et al.} \cite{alvarado2005} measured a KER of 36 eV for this channel in collisions of water molecules with ions. The calculations predicted by the PC model agree well with the experimental data, and the model can accurately simulate the KER of this channel. After obtaining the KER without nuclear reaction influence, we added the nuclear reaction. The simulation conditions are described as follows.
	
	I. Regarding the coordinates of each point charge, linear ammonia monohydrate was chosen for simulation [Fig. \ref{fig:1(a)}], with a distance of approximately 3.12 \text{\AA} between the N and O atoms \cite{Kollman1971}. The bond length of the water molecule measures 0.96 $\text{\AA}$, while the bond angle is 104.5$^{\circ}$ \cite{kim2016}. In this way, the position of each point charge can be obtained.
	
	\begin{figure}
		\centering
		\includegraphics[width=0.48\textwidth]{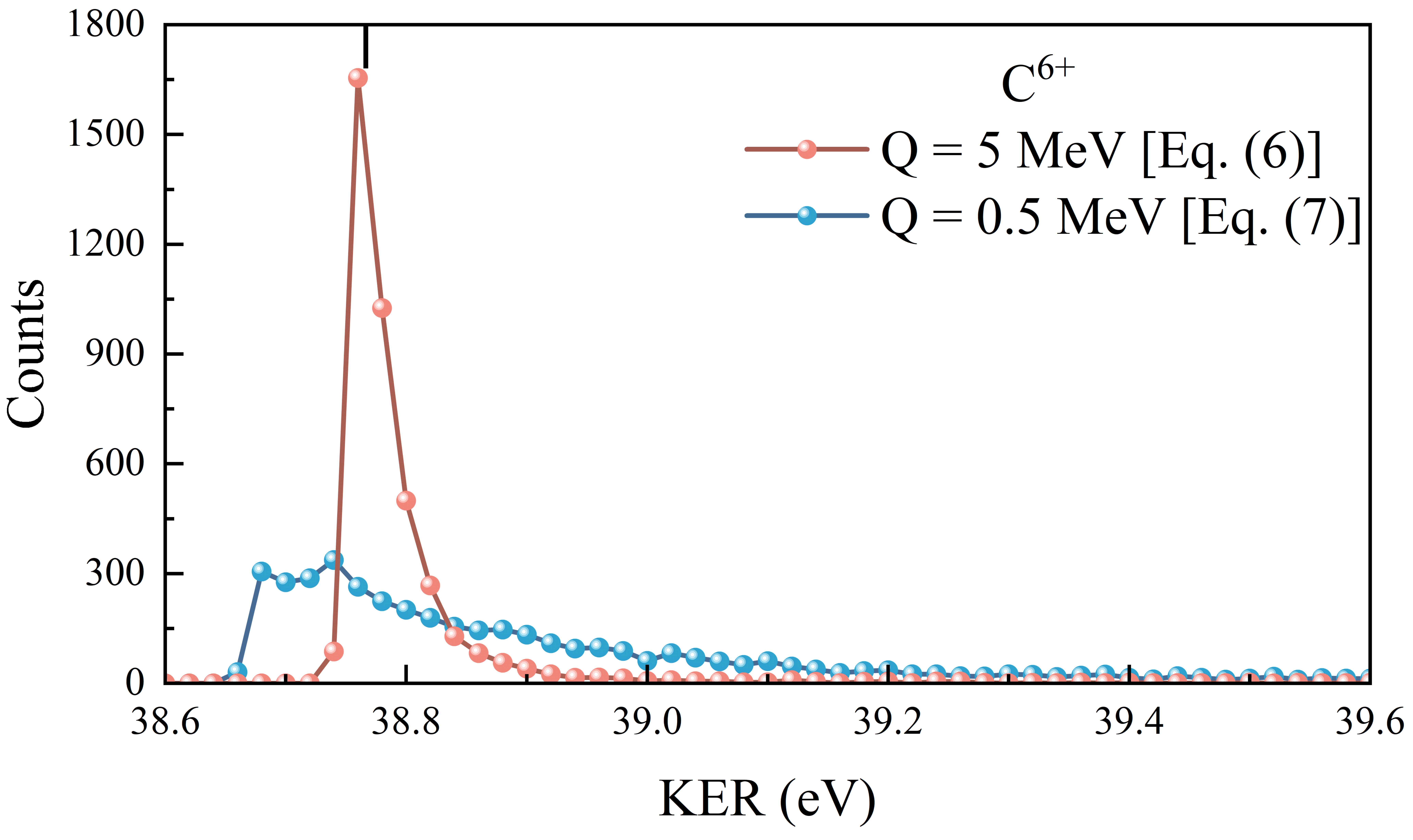}
		\vspace{-6mm} 
		\caption{\label{fig:3}KER distribution of water molecule fragments at nuclear-reaction energies $Q$ of 5 MeV and 0.5 MeV.}
		\vspace{-4.5mm}
	\end{figure}
	II. In setting the point charges, we refer to the nuclear reaction channel [Eq. (\ref{eq6})] and the fragmentation channel [Eq. (\ref{eq9})]. Two hydrogen atoms and one oxygen atom each have a positive charge.  $\alpha$ particle carries two positive charges. The charge of carbon ions may not be unique, and we model here three charge states, C$^{2+}$, C$^{4+}$, and C$^{6+}$.
	
	III. Regarding dynamics, it is assumed that two events occur at the same time: the decay of the compound nucleus $\rm{^{16}O}$ and the breaking of the chemical bond of the water molecule[Figs.\ref{fig:1(b)} and \ref{fig:1(c)}]. The nuclear-reaction energy $Q$ is completely transformed into the KE of the products, and the conservation of momentum and energy determines the velocities. $\alpha$ particle gets 3.75 MeV energy, and carbon ion is 1.25 MeV. The particles are emitted isotropically. The Coulomb force controls the subsequent motion process.
	
	Fig. \ref{fig:2} shows the KER distribution of water molecule fragmentation (hereafter referred to as KER) under the influence of the  ${\rm^{15}N(p,\alpha)^{12}C}$ reaction. The KER without nuclear reaction influence is 38.77 eV, which is referred to as the standard value indicated by the black dotted line. From Fig. \ref{fig:2(d)}, it can be observed that under the influence of nuclear reactions, over 95\% of KER values fall within the range of 38.75 to 38.9 eV. Carbon ions with the higher charge state exhibit a more pronounced KER broadening. The above phenomenon suggests that ${\rm^{15}N(p,\alpha)^{12}C}$ reactions lead to KER broadening, and the broadening is related to the charge state of the nuclear-reaction products. The blue dotted line represents the average KER under nuclear reaction influence. However, because extreme values with very small ratios significantly impact the mean, using the median (indicated by the green dotted line in Fig. \ref{fig:2}) is a better representative value. Comparison of the median with the standard values in Figs. \ref{fig:2(a)}, \ref{fig:2(b)}, and \ref{fig:2(c)}, the nuclear reactions lead to an overall increase in the KER, which is more pronounced with the increase in the charge state of the carbon ions. However, as shown in Fig. \ref{fig:2(c)}, the maximum difference between the median and the standard value is only 0.009 eV, and it is difficult to measure such a small effect in the experiment. As shown in Fig. \ref{fig:2(d)}, the ${\rm^{15}N(p,\alpha)^{12}C}$ reaction leads to a shift of the KER peak to lower energies.
	\begin{figure}
		\subfigure{
			\label{fig:4(a)}
			\includegraphics[width=0.46\textwidth]{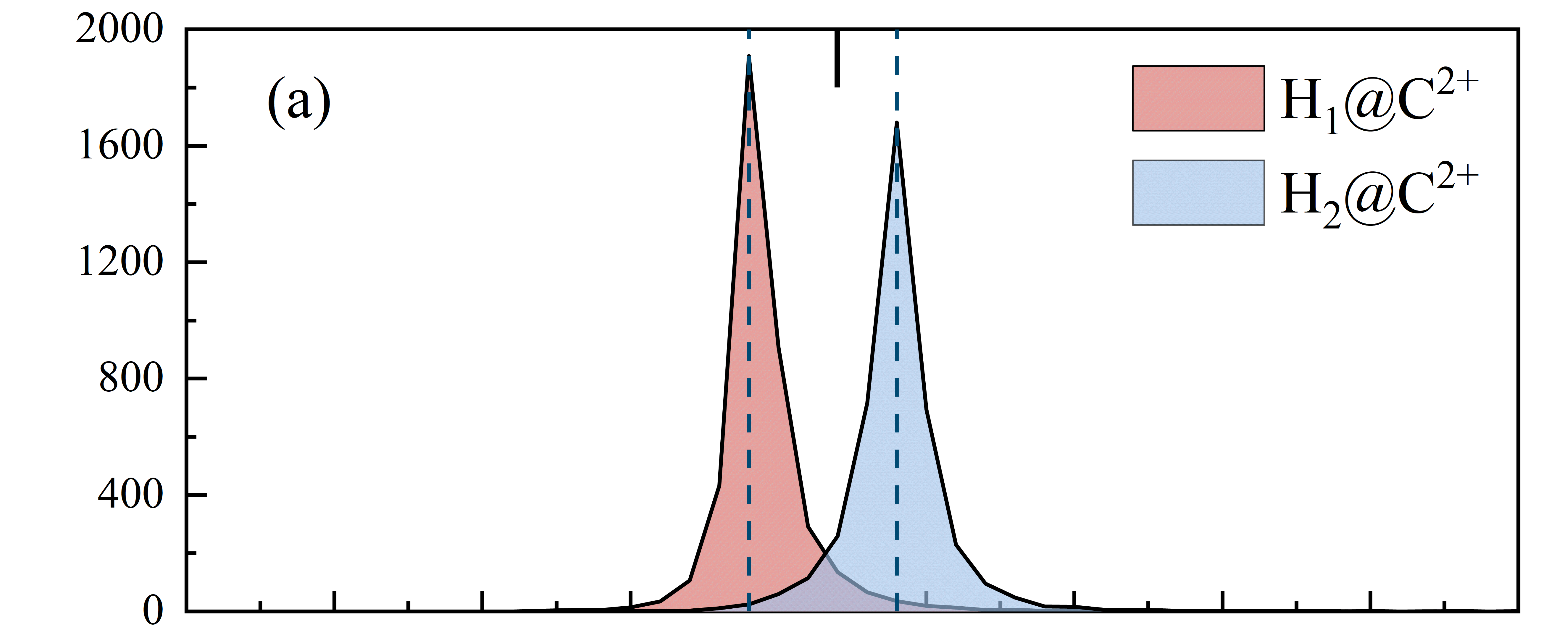}
		}
		\vspace{-3.85mm}
		
		\subfigure{
			\label{fig:4(b)}
			\includegraphics[width=0.46\textwidth]{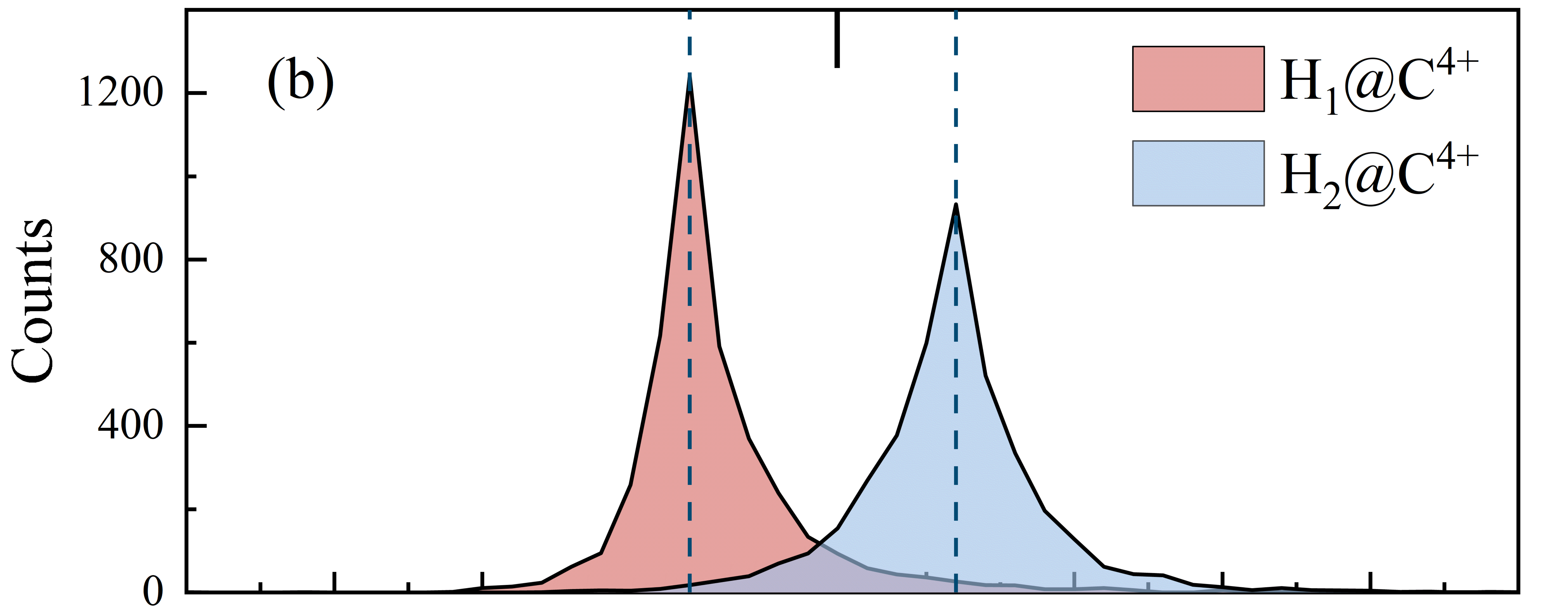}
		}
		\vspace{-3.85mm}
		
		\subfigure{
			\label{fig:4(c)}
			\includegraphics[width=0.46\textwidth]{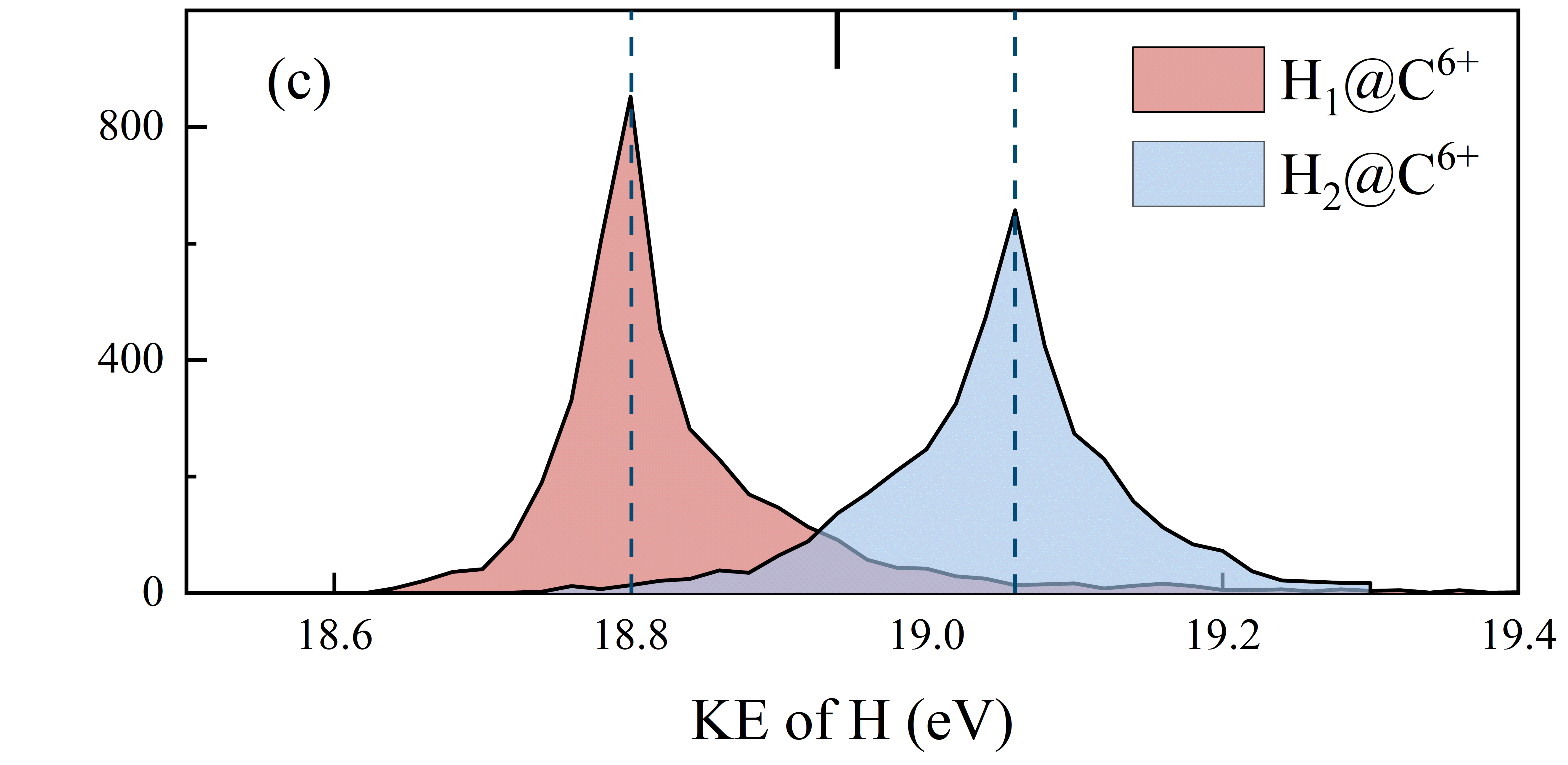}
		}
		\vspace{-2.5mm}
		
		\caption{\label{fig:4}Distribution of KE values of protons produced by water molecule fragmentation under the influence of nuclear reaction, the black vertical line represents the KE value of protons calculated by the PC model.}
		\vspace{-4mm}
		
	\end{figure}
	
	It was observed that changes in the velocity of nuclear-reaction particles significantly affect KER. Decreasing the KE of the nuclear-reaction products by an order of magnitude and simulated with C$^{6+}$, $q/v$ is approximately equal to $10$. The simulation results are shown in Fig. \ref{fig:3}. After decreasing the KE, the KER broadens significantly, with about 6\% of the values differing from the standard value (black vertical line) by more than 1 eV, a sizable broadening that can be measured experimentally with the COLTRIMS device. Thus, the KER is closely related to the values of $q$ and $v$ of the nuclear-reaction products. When the $q/v$ value becomes high, the KER width is systematically broadened and the peak position changes, which may lead to other effects.
	
	Figure \ref{fig:4} describes the KE distribution of the two proton fragments produced by the water molecule. H$_1$ represents protons closer to the nuclear reaction, while H$_2$ represents protons farther from the nuclear reaction. Due to the symmetric structure of the water molecule, the PC model calculates the same KE (represented by the black vertical line) for the two protons. However, nuclear reactions do not simply increase the KE of each fragment of the water molecule. For H$_1$, most KE values are lower than those computed by the PC model, while the opposite is observed for H$_2$. With the charge state of carbon ions increases, the difference in KE between the two protons becomes more significant as shown in Fig. \ref{fig:4(a)}, \ref{fig:4(b)} and \ref{fig:4(c)}. The KE distribution of each proton becomes broader. For the oxygen ion, the change trend is the same as H$_2$.
	
	In summary, simulations based on the point-charge approximation reveal the broadening effect of the ${\rm^{15}N(p,\alpha)^{12}C}$ reaction on the KER. The degree of broadening is affected by the $q/v$ values of the nuclear-reaction particles, i.e., higher charge states and lower velocities lead to a more pronounced broadening. When $q/v$ is approximately equal to 10, obvious changes in the KER are possible to be measured by the COLTRIMS. In addition, the study of the KE values of different fragments shows that the KE values of protons closer to the nuclear reaction (H$_1$) are lower than the standard values, and the opposite is true for protons farther away (H$_2$). Note that due to the low $Q$ value of reaction energy in some nuclear reaction channels, the KE obtained by nuclear-reaction particles is low. When the velocities of nuclear-reaction particles and molecular fragments are close to each other, it can greatly affect the KE of the fragments. It may lead to other unknown effects, such as higher energy deposition. Therefore, in a process similar to BNCT or PBCT, it is necessary to consider the effects of some channels with lower nuclear-reaction energy $Q$. In addition, charged particles from nuclear reactions ionize more water molecules, which may enhance biological effects. There are also other fragmentation channels of water molecules remain to be considered, which may have a more significant impact.
	
	    \vspace{3mm}
	This work was supported by the National Natural Science Foundation of China (Nos. 12374234 and 12074352), the Sichuan Science and Technology Program (Grant No. 2023ZYD0017), the Defense Industrial Technology Development Program (Grant No. JCKYS2023212808) and the Fundamental Research Funds for the Central Universities in China (Grant No. YJ202144).

	\bibliography{ref}

\end{document}